\documentclass[aps,twocolumn,showpacs,superscriptaddress,amsmath,uicr]{revtex4}
\usepackage[dvips]{graphicx}
\usepackage{amsmath}
\usepackage{amsfonts}
\usepackage{amssymb}
\usepackage{units}
\usepackage{dcolumn}
\usepackage{bm}
\usepackage[sort&compress]{natbib}
\bibpunct{[}{]}{,}{a}{}{;}

\newcommand{\be}{\begin{equation}}
\newcommand{\ee}{\end{equation}}
\newcommand{\bea}{\begin{eqnarray}}
\newcommand{\eea}{\end{eqnarray}}
\newcommand{\bean}{\begin{eqnarray*}}
\newcommand{\eean}{\end{eqnarray*}}

\begin{document}

\widetext

\title{Molecular hydrogen adsorbed on benzene: insights from a quantum Monte Carlo study}

\author{Todd D. Beaudet}
\email[]{beaudet@uiuc.edu}
\affiliation{ Department of Physics, University of Illinois at Urbana-Champaign,
1110 W. Green St, Urbana, IL 61801, USA}
\author{Michele Casula}
\email[]{casula@uiuc.edu}
\affiliation{ Department of Physics, University of Illinois at Urbana-Champaign,
1110 W. Green St, Urbana, IL 61801, USA}
\author{Jeongnim Kim}
\email[]{jnkim@uiuc.edu}
\affiliation{National Center for Supercomputing Applications, 1205 W. Clark St., Urbana, IL 61801, USA}
\affiliation{Materials Computational Center, University of Illinois at Urbana-Champaign, 104 S. Goodwin Ave., Urbana, IL 61801, USA}
\author{Sandro Sorella}
\email[]{sorella@sissa.it}
\affiliation{ SISSA, International School for Advanced Studies,
  34014, Trieste, Italy}
\affiliation{ DEMOCRITOS, National Simulation Center,
  34014, Trieste, Italy}
\author{Richard M. Martin}
\email[]{rmartin@uiuc.edu}
\affiliation{ Department of Physics, University of Illinois at Urbana-Champaign,
1110 W. Green St, Urbana, IL 61801, USA}

\date{\today}
\bibliographystyle{h-physrev3}

\begin{abstract}
We present a quantum Monte Carlo study of the hydrogen-benzene system where binding
is very weak. We demonstrate that the binding is well described at both variational Monte Carlo (VMC)
and diffusion Monte Carlo (DMC) levels by a Jastrow correlated single determinant geminal
wave function with an optimized compact basis set that includes diffuse orbitals.
Agreement between VMC and fixed-node DMC binding energies is found
to be within 0.18 mHa, suggesting the calculations are well-converged with respect to the basis.
Essentially the same binding is also found in independent DMC calculations using a different
trial wave function of a more conventional Slater-Jastrow form, supporting our conclusion that the binding energy
is accurate and includes  all effects of correlation. We compare with empirical models and previous calculations,
and we discuss the physical mechanisms of the interaction, the role of diffuse basis functions,
and the charge redistribution in the bond.
\end{abstract}


\maketitle

\section{Introduction}
\label{introduction}

A great deal of research has gone into the study of hydrogen storage materials
due in large part to the prospect of zero emissions transportation. The value of
finding a material that effectively and reversibly stores hydrogen can hardly be
overstated.\cite{TowardAHydrogenEconomy2004}
Hydrogen storage has a long history starting with the discovery of reversible
hydriding in palladium in 1866 by Graham,\cite{Graham1866} with much work
on systems such as metal hydrides\cite{Wiswall1978} and complex
hydrides.\cite{CXLLT2002,miwa:195109,magyari-kope:220101,araujo:165101}
Recently, materials such as carbon nanotubes,\cite{dillon:1997}
fullerenes,\cite{zhao:155504}
metal-organic-frameworks (MOF),\cite{Rosi05162003,LochanReview2006,Fichtner2005}
and others have been studied.
The U.S. Department of Energy has set a goal for 2010
to find a material that can meet the requirements of
storing molecular hydrogen at 6\% weight and 45 grams per liter,
which is nearly half the hydrogen content of water, reversibly
in the range of -30--50$^\circ$C for a thousand cycles.\cite{www1.eere...mypp2007}
While many solutions to this problem have been
offered, none have satisfied all these constraints.

In this paper we study the 
adsorption of molecular hydrogen on a benzene ring as a test system. 
By itself this system is not expected to be a practical system for storage 
due the expected binding energy being much weaker than the 
target range of 20--40 kJ/mol H$_2$ ($\sim$ 7--15 mHa/H$_2$) needed for reversibility 
in the desired temperature range.\cite{LochanReview2006} 
However, benzene is similar to the 5- or 6-member rings that are
characteristic building blocks of all the carbon systems mentioned above. 
As such, an accurate description of this structure is highly relevant to ongoing research
where hydrogen is adsorbed on or around carbon rings.
Besides this, the hydrogen-benzene system is a good test case for theoretical predictions 
because of the stringent requirements to reliably determine binding energies at 
the desired accuracy level. 
Further, a careful study of this system 
is an important test of the transferability of empirical potentials\cite{mattera_1980,crowell_1982}
that have been constructed primarily from experimental data on graphitic systems. 
In this paper we consider only the case where the hydrogen dimer is oriented
along the $C_6$ symmetry axis of the benzene molecule.
Other papers\cite{HamelCote2004,Hubner2004} have found this to
be the favored configuration and orientational differences are not taken into account
in this work, since our main purpose is to present benchmark
calculations for the most stable geometry.

There have been many previous studies of the binding of H$_2$ on benzene
and related systems using various methods including density functional theory
(DFT), M\o ller-Plesset second order perturbation theory (MP2),
coupled cluster (CC) with single and double excitations (CCSD), and variations
of these.\cite{HamelCote2004,Hubner2004,LochanReview2006}
The values obtained so far for the binding energy, falling in the range of $0.4$--$1.9$ mHa,
\cite{HamelCote2004,Hubner2004,LochanReview2006}, are very small
and require a high level of accuracy of all the methods.
The DFT calculations have great advantage as they are fast, scale well
with system size, and can be readily converged with respect to the basis.
However, the accuracy of their results is limited
by the approximation on the exchange and correlation functionals, and
there is no known way to systematically improve it. The many-body CC methods are
the most accurate, although their applications are limited to small systems and not-so-large
basis set due to poor scaling with the number of electrons
and the size of the basis. Perturbation methods such
as MP2 are valuable theories, with a system size scaling better than any CC method,
but with intermediate accuracy. In the H$_2$-benzene system,
perhaps the most accurate results to date have been derived from MP2 and CCSD(T) calculations 
by H\"{u}bner \textit{et al.}\cite{Hubner2004} 
They found that the binding increases with 
increasing basis size in MP2 calculations, whereas it decreases as the level of the theory is improved to 
CCSD(T). Based on the best MP2 binding energy (1.87 mHa) and the best CCSD(T) value (1.16 mHa) with 
affordable bases, they estimated the actual binding energy to be $\sim 1.5$ mHa. 
It should be noted that these numbers have already included 
basis set superposition error corrections as high as 0.36 mHa 
and so represent a substantial fraction of the binding energy. 
Such methods necessitate carefully extrapolating the results with respect to the basis 
set and the level of theory. However, such extrapolations represent a difficulty due to the high computational 
cost of large bases, particularly in the CCSD(T) framework. 

In the present work we study the hydrogen-benzene problem using
quantum Monte Carlo (QMC) methods,
that offer several advantages: many-body correlation effects can be explicitly included in the wave
function, scaling with the system size is favorable like DFT,
and calculations are variational and usually less dependent on the basis set.
For a review and references to earlier work, see Ref.~\onlinecite{FoulkesReview2001}.
By means of QMC,
a trial correlated wave function can be optimized in the
variational Monte Carlo (VMC) framework,\cite{umrigar:110201,sorella:014105}
and its energy can be further minimized by the diffusion Monte
Carlo (DMC) algorithm, which stochastically projects the
optimized VMC (trial) wave function to the ground state.
The only fundamental limitation is the  well known ``sign problem''  for fermion systems, that does not allow a numerically stable calculation.
Therefore,  in this case, 
the so called fixed node  (FN) approximation is adopted
by constraining the diffusion within the nodal pockets of the initial variational wave
function.\cite{FoulkesReview2001} 
Thus, the FN DMC method is unbiased only if the nodes
of the trial wave function coincide with those of the true ground state.
We addressed the issue of the FN bias in two ways.
First, we used advanced QMC optimization methods\cite{sorella:014105,sorella:241103,umrigar:150201,umrigar:110201}
and physical principles to find a Jastrow correlated antisymmetrized geminal product (JAGP)\cite{casula:7110,casula:6500}
which gives a VMC binding energy with an accuracy comparable to the post Hartree-Fock (HF) methods.
In addition, we computed the binding energy at the DMC level
using the JAGP and a simpler Slater-Jastrow (SJ) form with a 
PBE-DFT optimized basis set as the trial wave function. 
The agreement found between them supports clearly the idea that our results are independent
of the basis set and variational form, and it is a check
for the accuracy of our DMC calculations against the FN approximation,
since the nodes of the two wave functions are {\it a priori} different.
This is encouraging for another reason:
although the SJ trial function is not as accurate as
the JAGP at the VMC level, it is more easily extended to larger systems
which are important for future work.
The necessary condition for that agreement is using a basis sufficiently extended in the tails.
This is not surprising for a system driven by Van der Waals (VdW)
interactions which lead to weak binding and large equilibrium distance,
as pointed out in Ref.~\onlinecite{diedrich:184106},
but it is a crucial point since the tails are not very important in the total energy.
Our work shows that DMC can capture the correct binding as long as the basis is extended
enough to allow accurate sampling of the outer regions of the molecules.
This is brought out by a detailed study of the electron density changes due to binding.

The paper is organized as follows. In Sec.~\ref{computational_details}
we describe the QMC methods employed
as well as the SJ and JAGP wave functions that serve as the variational guess
in our QMC calculations.
In Sec.~\ref{results} we discuss our results on the binding energy of the hydrogen-benzene system,
where the hydrogen is oriented perpendicular to and centered over the benzene at various molecular
spacings. In Sec.~\ref{otherwork} we compare our findings to previous works.
In Sec.~\ref{discussion} we discuss the physics of the hydrogen-benzene bond in terms of its electron density,
by comparing the QMC and DFT-PBE results. Finally, we draw our conclusions in Sec.~\ref{summary}.

\section{Computational details}
\label{computational_details}

The unique feature of QMC methods is that they involve directly the
many-body wave function and address the necessary high-dimensional integrals
by stochastically sampling the configurational space.\cite{PhysRev.138.A442,FoulkesReview2001}
In so doing, the wave function
may take on a form more general
than a bare linear combination of Slater determinants
while still maintaining computational efficiency. In the following Subsections
we describe the variational
trial wave function (Subsec.~\ref{wave_functions}), and
provide a brief introduction on the methodology (Subsec.~\ref{methods}).

\subsection{Wave functions}
\label{wave_functions}

A wave function that describes a system of $N$ identical fermions must be antisymmetric under particle exchange.
To simplify the description of such a wave function, it is often useful to factor the
wave function into a positive symmetric part, called the Jastrow factor, and an antisymmetric part
so that a wave function can be expressed as
\begin{equation}\label{eq:slaterjastrow}
\Psi(\mathbf{x}_{1},\ldots,\mathbf{x}_{N}) = J(\mathbf{x}_{1},\ldots,\mathbf{x}_{N})\,\Psi_{AS}(\mathbf{x}_{1},\ldots,\mathbf{x}_{N}).
\end{equation}
where $\mathbf{x}_i\equiv\{\mathbf{r}_i,\sigma_i\}$ is a space-spin coordinate,
$J(\mathbf{x}_{1},\ldots,\mathbf{x}_{N})$ is the Jastrow factor,
and $\Psi_{AS}(\mathbf{x}_{1},\ldots,\mathbf{x}_{N})$ is the antisymmetric part.
The Jastrow can be further factored into one-body, two-body, three-body,
and higher-body terms ($J=J_1J_2J_3\cdots$) which correspond to effective electron-ion,
electron-electron, electron-electron-ion, etc. interactions.

One of the choices of trial function is to approximate the
antisymmetric wave function as a single Slater determinant
of spin orbitals. If there are no spin orbit interactions, the
energy depends only upon the spatial part of the wave function 
which can be written as a product of spin up and spin down determinants.
In the unpolarized case, the spatial form is given by 
\begin{widetext}
\begin{equation}\label{eq:spinfactored}
\Psi_{AS}(\mathbf{x}_{1},\ldots,\mathbf{x}_{N}) = \left\vert
\begin{array}{ccc}
    \varphi_1(\mathbf{r}_1^\uparrow)                 & \ldots & \varphi_{\nicefrac{N}{2}}(\mathbf{r}_1^\uparrow) \\
    \vdots                                           & \ddots & \vdots \\
    \varphi_1(\mathbf{r}_{\nicefrac{N}{2}}^\uparrow) & \ldots & \varphi_{\nicefrac{N}{2}}(\mathbf{r}_{\nicefrac{N}{2}}^\uparrow)
\end{array} \right\vert \left\vert
\begin{array}{ccc}
    \varphi_{\nicefrac{N}{2}+1}(\mathbf{r}_1^\downarrow)                 & \ldots & \varphi_N(\mathbf{r}_1^\downarrow) \\
    \vdots                                                               & \ddots & \vdots \\
    \varphi_{\nicefrac{N}{2}+1}(\mathbf{r}_{\nicefrac{N}{2}}^\downarrow) & \ldots & \varphi_N(\mathbf{r}_{\nicefrac{N}{2}}^\downarrow)
\end{array} \right\vert,
\end{equation}
\end{widetext}
where each $\varphi_i(\mathbf{r})$ is a single-body space orbital and N is the total number of electrons. 
In our calculations the single-body orbitals $\varphi_i(\mathbf{r})$ are derived
using the Perdew-Burke-Ernzerhof\cite{PhysRevLett.77.3865,PhysRevLett.78.1396} (PBE) density functional 
with the VTZ Gaussian basis~\cite{burkatzki:234105} modified to include diffuse functions from the
aug-cc-pVTZ basis.\cite{kendall:6796} The normalization factor in Eq.~\ref{eq:spinfactored} has been
dropped since in the QMC approach the wave function does not need to be normalized.

Since the Slater determinant description alone is unable to treat correlation effects,
the Jastrow factor is used to include electron correlation in a computationally efficient way.
Moreover, it provides a way to enforce cusp (or coalescence) conditions
which are an exact property of the many-body wave function, and demand
that the local energy $E_L\equiv\Psi^{-1}H\Psi$ remains finite as electrons approach ion centers and each other.
The Jastrow factor we applied to the Slater determinant is a Wagner-Mitas form \cite{wagner:034105}
modified so that the electron-ion and electron-electron cusp conditions are fulfilled.
The one- and two-body Jastrow terms are given by
\begin{equation}\label{eq:wmjastrow1}
  J_{1}(\mathbf{R}) = \prod_{ia} \exp\left[\sum_{k} (b_{ak}r_{ia} + c_{ak}) \upsilon_{ak}(r_{ia})\right]
\end{equation}
and
\begin{equation}\label{eq:wmjastrow2}
  J_{2}(\mathbf{R}) = \prod_{i<j} \exp\left[\sum_{k}(b_{k}r_{ij} + c_{k}) \upsilon_{k}(r_{ij})\right]
\end{equation}
where $\mathbf{R}=\{\mathbf{r}_{1},\ldots,\mathbf{r}_{N}\}$ specifies the $N$ electron space coordinates,
$i$ and $a$ index electrons and nuclei respectively, $r_{ia}$ and $r_{ij}$ are electron-ion
and electron-electron distances, and $k$ indexes the expansion terms.
In our work we used three terms and, when needed, one cusp term.
In the above Equations, $\upsilon_{k}(r) = (1 - z(r/r_{cut}))/(1+\beta_{k}z(r/r_{cut}))$, with $z(x)=x^{2}(6-8x+3x^{2})$ and
parameters $b, c, \beta$ optimizable (with the exception of those that are cusp dependent). The function
$z(x)$ has the properties $z(0)=z'(0)=z'(1)=0$ and $z(1)=1$, so that the Jastrow has a well defined cutoff at
$r_{cut}=10$ Bohr. Cusps between same spin electrons are not accounted for. This is justified because
of the Pauli exclusion principle, which keeps them apart. Also the three-body terms have been neglected.
It should be emphasized that the single-body Slater orbitals 
obtained from PBE-DFT are not further optimized since we would like to check the accuracy
of the PBE-DFT nodes with respect to a more correlated and fully optimized wave function, such as the JAGP
form described below.
However, optimizing the above Jastrow is convenient as it improves the VMC energy and variance 
and shortens the DMC projection time, without changing the nodes.

The other trial function used in this work is the JAGP, where
the antisymmetric part is a single determinant of two-body orbitals (geminals).
This approach has been successfully applied in several contexts where electron correlations play a significant role.
For example, the JAGP form is related to the pairing in the BCS wave function for superconductivity,\cite{PhysRev.106.162,super}
the resonating valence bond (RVB) proposed by Pauling in 1939,\cite{Pauling1960}
and can be used to describe strongly-correlated electrons in transition metals.
Recent applications in quantum chemistry include benzene,\cite{casula:7110} 
the benzene dimer interacting via weak van der Waals forces,\cite{sorella:014105} 
and iron dimer.\cite{casula:unpublished2007} 

Since the ground state of the hydrogen-benzene system is an unpolarized spin singlet ($N^\uparrow = N^\downarrow = N/2$)
the spatial part of the AGP wave function can be written as a determinant of pairing functions\cite{bouchaud:553}
without including unpaired orbitals, namely

\begin{equation}\label{eq:JAGPdeterminant}
\Psi_{AS}(\mathbf{X}) =
\left\vert
\begin{array}{ccc}
    \phi(\mathbf{r}_1^\uparrow,\mathbf{r}_1^\downarrow) & \ldots
    & \phi(\mathbf{r}_1^\uparrow,\mathbf{r}_{\nicefrac{N}{2}}^\downarrow) \\
    \vdots & \ddots & \vdots \\
    \phi(\mathbf{r}_{\nicefrac{N}{2}}^\uparrow,\mathbf{r}_1^\downarrow) & \ldots
    & \phi(\mathbf{r}_{\nicefrac{N}{2}}^\uparrow,\mathbf{r}_{\nicefrac{N}{2}}^\downarrow)
\end{array}
\right\vert,
\end{equation}
where $\mathbf{X}=\{\mathbf{x}_{1},\ldots,\mathbf{x}_{N}\}$ specifies the $N$ electron space-spin coordinates
and the paring function $\phi(\mathbf{r}_i^\uparrow,\mathbf{r}_j^\downarrow)$ can be expanded in single-body atomic
orbitals so that
\begin{equation}\label{eq:pairingfunction}
\phi(\mathbf{r}_i^\uparrow,\mathbf{r}_j^\downarrow)=\sum_{lmab}\lambda_{ab}^{lm}
\varphi_{al}(\mathbf{r}_{i}^\uparrow)\varphi_{bm}(\mathbf{r}_{j}^\downarrow)
\end{equation}
where $l$ and $m$ index the orbitals centered on ions $a$ and $b$ respectively.
Here also, as in the SJ wave function, Gaussian type orbitals are used.

The Jastrow factor in the JAGP wave function is somewhat different from the one applied to the Slater determinant.
The cusp conditions are fulfilled through the one-body $J_1$, and two-body $J_2$ Jastrow terms, written as
\begin{equation}\label{eq:JastrowOneBody}
J_1(\mathbf{R}) = \prod_{ia}\exp\left[-(2Z_a)^{\nicefrac{3}{4}}u((2Z_a)^{\nicefrac{1}{4}}r_{ia})\right],
\end{equation}
and
\begin{equation}\label{eq:JastrowTwoBody}
J_2(\mathbf{R}) = \prod_{i<j}\exp\left[u(r_{ij}))\right],
\end{equation}
where $\mathbf{R}$ is an all-electron configuration, $i$ and $j$ are electron indices, and
$a$ is a nuclear index. The ion centers have effective charge $Z_a$ and the
function $u(x)$ satisfies the electron-ion and electron-electron cusp conditions between unlike-spin particles
with $u(0)=0$ and $u^\prime(0)=\frac{1}{2}$.
Here, $u(r)\equiv\frac{F}{2}\left(1-e^{-\nicefrac{r}{F}}\right)$, where $F$ is an optimizable parameter.
In Eq.~\ref{eq:JastrowOneBody}, the argument of $u$ is multiplied by $(2Z_a)^{\nicefrac{1}{4}}$ in order to satisfy the 
random phase approximation behavior at large $r_{ia}$.\cite{markus} 

A distinguishing feature of the JAGP with respect to the simple SJ wave function is the presence of
electron-electron-ion and electron-ion-electron-ion terms,
conventionally referred to as three- and four-body Jastrow factors. In the JAGP wave function,
they are written as the exponential of a pairing function like the one in Eq.~\ref{eq:pairingfunction}, namely
\begin{equation}\label{eq:JastrowThreeBody}
J_{34}(\mathbf{R}) = \prod_{ij}\exp
\left[-\sum_{ablm} g_{lm}^{ab}\chi_{al}(\mathbf{r}_i^\uparrow)\chi_{bm}(\mathbf{r}_j^\downarrow) \right].
\end{equation}
Here $g_{lm}^{ab}$ are optimizable parameters and $l$ ($m$) is an index for single-particle Gaussian orbitals $\chi_{al}$
centered on nucleus $a$ ($b$).
The three- and four-body Jastrow terms provide for electron-correlations substantially beyond the largely
cusp related one- and two-body terms and are able to describe subtle effects like van der Waals forces.\cite{casulaPhDthesis:2005}
However, Eq.~\ref{eq:JastrowThreeBody} does not include the three-body cusp conditions
recently derived by Fournias \emph{et al.},\cite{fournais} which can improve the quality of the nodes
of the JAGP wave function described here.
The effect of the three-body cusp conditions in the energy optimization
and nodal structure is presently under investigation.

\subsection{Methods}
\label{methods}

In setting up our Hamiltonian, we use the Born-Oppeheimer approximation, a Hartree-Fock norm conserving soft
pseudopotential for the He core of carbon
\footnote{The pseudopotentials we used are norm-conserving Hartree-Fock generated by 
E. Shirley's code
with the construction by D. Vanderbilt, Phys. Rev. B \textbf{32}, 8412 (1985).}, 
and the bare Coulomb potential for hydrogen and electron-electron interactions.
Our procedure is to start with a trial wave function
which includes variational parameters (see Subsec.~\ref{wave_functions} for the forms employed in this work).
We proceed to optimize its energy and variance
at the VMC level using minimization methods suitable for the particular
form.\cite{HestenesStiefel1952,PhysRevB.64.024512,sorella:014105,umrigar:110201}
The resulting analytic wave function is projected to the FN ground state using
DMC methods\cite{casula:100201,casula:161102} 
recently developed to yield a stable simulation and an upper bound of the ground state energy
even for non-local pseudopotentials.

As we mentioned above, we use the full electron-nucleus Hamiltonian except for the carbon core which
is replaced by a pseudopotential.
This leads to a better statistics due to a narrower energy scale,
a reduction in the number of optimization parameters,
a more stable optimization of our JAGP wave function,\cite{sorella:014105}
and a larger DMC time step needed for convergence, which results in a cheaper computational
cost of the simulation.
On the other hand, its drawback is that part of the fully local Coulomb potential
is replaced by a non-local pseudopotential $V_{\textrm{non-local}}$ that is angular
momentum dependent. Within the  VMC framework the corresponding angular integration of the non-local 
potential remains  possible since the wave function is known analytically. However, problems arise
in the FN DMC because the FN ground state is given only by a stochastic 
sampling. A partial solution is the localization approximation (LA),
where the trial (or guiding) wave function $\Psi_G$
is used to approximate the projected ground state so that the non-local pseudopotential
terms can be evaluated.\cite{FoulkesReview2001} 

This changes the Hamiltonian so that the projected energy is no longer a variational
upper bound of the original non-local FN Hamiltonian.
The FN Green's function positive definite character becomes impaired by the very
attractive parts of the non-local pseudopotential which has been made local using the guiding wave function.
Indeed, on the nodes of the guidance the localized potential
can diverge negatively leading to possible numerical instabilities where the walker
population can grow without bound. Our method of implementing non-local pseudopotentials in the FN DMC methods
sidesteps these problems entirely.

Our FN DMC calculations are done with either continuous or lattice regularized (LRDMC) moves both of which
utilize a common means of addressing the inherent problems of the LA.
In contrast to the LA, we use a breakup\cite{casula:100201,casula:161102}
of the non-local potential that localizes the positive matrix
elements into the branching term while treating the negative matrix elements
as a non-local diffusion operator sampled via a \emph{heat bath} scheme.\cite{casula:161102}
The positive and negative terms are defined by
\begin{equation}\label{eq:nonlocalbreakup}
V^\pm_{\mathbf{R}^\prime,\mathbf{R}}=1/2(V_{\mathbf{R}^\prime,\mathbf{R}}\pm |V_{\mathbf{R}^\prime,\mathbf{R}}|)
\end{equation}
where
\begin{equation}\label{eq:nonlocal}
V_{\mathbf{R}^\prime,\mathbf{R}}=\frac{\Psi_G(\mathbf{R}^\prime)}{\Psi_G(\mathbf{R})}
\left<\mathbf{R}^\prime|V_{\textrm{non-local}}|\mathbf{R}\right>,
\end{equation}
and $\mathbf{R}$, $\mathbf{R}'$ are all-electron configurations on a quadrature mesh
with one electron rotated around a pseudo ion.\cite{fahy}
The breakup corresponds to 
an effective Hamiltonian $H^{\textrm{eff}}$, defined as
\begin{eqnarray}
\label{H_effective}
H^{\textrm{eff}}_{\mathbf{R},\mathbf{R}} & = & K + V^{\textrm{eff}}(\mathbf{R}) \\
H^{\textrm{eff}}_{\mathbf{R}^\prime,\mathbf{R}} & = & \left<\mathbf{R}^\prime|V_{\textrm{non-local}}|\mathbf{R}\right>
\qquad \textrm{if $V_{\mathbf{R}^\prime,\mathbf{R}}<0$},\nonumber
\end{eqnarray}
with the modified local potential $V^{\textrm{eff}}(\mathbf{R})=
V_{\mathrm{loc}}(\mathbf{R}) + \sum_{\mathbf{R}^\prime} V^+_{\mathbf{R}^\prime,\mathbf{R}}$
that includes the \emph{sign flip} terms.
The FN ground state energy of the Hamiltonian in Eq.~\ref{H_effective}
is a variational upper bound of the original non-local Hamiltonian.\cite{dutch} 
Furthermore, DMC stability is improved substantially due to a softening of the most attractive parts of the localized
pseudopotential. In the LA, the highly attractive regions of the localization can result in a walker population ``blow up''
so that the calculation ends suddenly. Moving the negative
part of the localization into a diffusion-like term causes the walkers to be driven away from such regions.

The main difference between the DMC Hamiltonian reported in Eq.~\ref{H_effective}
and the LRDMC is the kinetic operator $K$, which is replaced by a discretized 
$K^a$ in the LRDMC approach, and treated on the same footing as $V_{\textrm{non-local}}$.
$K^a$ is a linear combination of two discrete operators with incommensurate lattice spaces $a$ and $a'$
($a' = \nu a$, with $\nu$ an irrational number $>1$), namely
\begin{equation}
\label{k_operator}
K^a= -\frac{\eta}{2} ( \Delta^{a,p} + \Delta^{a',1-p} ),
\end{equation}
where $\Delta^{a,p}$ is the discretized Laplacian with mesh $a$ and weighting function $p$
(see Refs.~\onlinecite{casula:100201} and \onlinecite{sorella:014105}), and $\eta=1+ \mu a^2$
is a prefactor with the parameter $\mu$ tunable to improve the efficiency of the diffusion process.
Working with two incommensurate meshes helps to sample densely the continuous space by
performing discrete moves of length $a$ and $a'$.
The finest hop samples more likely
regions near atomic centers while the coarser one samples more often valence regions, the
result being an efficient sampling of the overall configuration space.
The difference between the continuous
and discretized local kinetic energies is added to $V^{\textrm{eff}}(\mathbf{R})$,
resulting in a mesh dependent potential
\begin{equation}\label{eq:discretizedpotential}
V^a(\mathbf{R})=V^{\textrm{eff}}(\mathbf{R})+\left[\frac{(K-K^a)\Psi_G}{\Psi_G}\right](\mathbf{R}).
\end{equation}
The consequence is a faster convergence of the energies in the $a\rightarrow 0$ extrapolation.
In spite of the discretization of $K$ (Eq.~\ref{k_operator})
and the redefinition of $V^{\textrm{eff}}$ (Eq.~\ref{eq:discretizedpotential}),
the LRDMC method is equivalent to the continuous space FN DMC with Hamiltonian in Eq.~\ref{H_effective}.
Indeed, in the limit of small mesh sizes $a$ and $a^\prime$,
the discretized Hamiltonian $H^a$ approaches the continuous $H$.
The usual DMC Trotter breakup results in a time step error
while the LRDMC paradigm results in a space step error, but both
share the same upper bound property in the zero-time-step zero-lattice-space limit
and converge to the same projected FN energy.\cite{casula:161102}

Our SJ calculations were done using continuous space DMC with QMCPACK.\cite{qmcpack} 
This code provides many features that make it easy to work with SJ wave functions.
The LRDMC method, available in the TurboRVB,\cite{TurboRVB} has been applied 
to the JAGP wave function after a full optimization of its parameters. We used two optimization procedures.
For the SJ work we employed the method of conjugate gradients (CG) introduced
by Hestenes and Stiefel\cite{HestenesStiefel1952} in 1952. This is a first-derivative method that finds the minimum of
a cost function (in our case a linear combination of the variance and the
energy), in a number of steps significantly smaller than the standard
steepest descent method, because for a quadratic cost function
it converges in a finite number of
iterations,  at most equal to the dimension of the vector space.\cite{PayneReview1992,Martin2004}
We optimized 10 parameters of the Jastrow functions but used the same VTZ basis set at all separations.
While systems are generally not quadratic, they are so near stationary points making the CG approach widely
applicable. Besides this, the method uses little memory, each iteration is equal time, and, while not guaranteed, the global minimum
is usually found. However, the statistical noise inherent in the QMC framework limits the applicability of our CG implementation to
systems involving not too many parameters, such as our SJ optimization.

The JAGP optimization, on the other hand, involves
a large number ($\sim1000$) of parameters,
mainly coming from the $\lambda_{ab}^{lm}$ (Eq.~\ref{eq:pairingfunction})
and $g_{lm}^{ab}$ (Eq.~\ref{eq:JastrowThreeBody}) matrices
in the AGP and Jastrow geminal expansions over the atomic basis set.
Therefore, an optimization technique robust under stochastic conditions is required.
The stochastic reconfiguration (SR) method recently
introduced by one of us (S.S.) \cite{PhysRevB.64.024512} in conjunction with subsequent
improvements\cite{sorella:014105,sorella:241103,umrigar:150201,umrigar:110201} has
been shown to be very efficient 
to minimize the variational energy.
This is a first-derivative
algorithm that moves the parameters $\alpha_k$ at each iteration according to
the update
\begin{equation}\label{eq:srmoves}
\alpha_k^\prime = \alpha_k + \delta\alpha_k
\end{equation}
where
\begin{equation}\label{eq:srdirection}
\delta\alpha_k = \gamma \sum_{k^\prime} s^{-1}_{k,k^\prime}\;f_{k^\prime}.
\end{equation}
Here, $k$ indexes the parameters, $f_k \equiv -\partial E/\partial\alpha_k$ are the generalized forces,
$s_{k,k^\prime}$ are the SR matrix elements, and $\gamma$ scales the length of the move.
The SR matrix is only required to be positive definite in order to lower the energy.
However, to make the convergence more efficient,
we define the SR matrix as\cite{PhysRevB.64.024512,casula:6500,casula:7110,sorella:014105}
\begin{equation}\label{eq:srmatrix}
s_{k,k^\prime} = \left(1+\delta_{k,k^\prime}\;\epsilon\right) \left(\left<O_k O_{k^\prime}\right> -
                \left<O_k\right>\left<O_{k^\prime}\right>\right)
\end{equation}
where $O_k=\partial_{\alpha_k}\ln |\left<\mathbf{R}|\Psi_G\right>|$ and $\epsilon$ is a cutoff chosen large
enough to guarantee the SR matrix remains well conditioned. While the SR matrix is guaranteed to be positive definite
even for a finite Monte Carlo sample, too small eigenvalues can result
in an amplification of errors coming from the forces $f_k$ through Eq.~\ref{eq:srdirection}.
Setting $\epsilon$ to a finite but small value
(usually $\simeq 10^{-3}$ smaller than the largest eigenvalue) makes
the optimization much more stable.\cite{sorella:014105}
The $\gamma$ factor in Eq.~\ref{eq:srdirection}
can be tuned  to speed up the convergence 
of standard SR 
by allowing for a better estimation of the magnitude of the parameter 
changes $\delta \alpha_k$. 
This is done
by evaluating the Hessian along the $\textbf{s}^{-1}\textbf{f}$ direction and finding
the minimum by assuming quadratic curvature. 
We extend this idea further and minimize the energy by
evaluating the Hessian at each step over the latest $n$ SR directions, with $n$ chosen to maximize the efficiency,
as explained in Ref.~\onlinecite{sorella:014105}.

\section{Results}
\label{results}

In this section we present results for
hydrogen-benzene binding where the hydrogen molecule
is oriented along the $C_6$ symmetry axis of the benzene molecule.
Previous studies\cite{HamelCote2004,Hubner2004} found this configuration
the most stable.
Here, we do not take into account other possible orientations, because our goal is to check the accuracy
of different QMC wave functions and provide benchmarks for the lowest energy configuration.
In order to resolve its potential energy surface, we consider the system at different
molecular center-of-mass separations $R$.
In our QMC calculations we have kept the geometry of each molecule fixed and
close to its experimental structure.\footnote{The actual bond lengths used in this work are:
$C-C = 2.63$ Bohr, $H-C = 2.04$ Bohr, and $H-H = 1.40$ Bohr. They are close to the best experimental and theoretical
values.\cite{feller,huber}}
We checked the effect of relaxing the geometries at the MP2 level and found
an energy lowering on the order of $\mu$Ha, whose effect is completely negligible in this case.

We emphasize that all our QMC results do not involve any corrections and are direct
energy differences with the largest computed distance ($R=15$) taken as reference for the zero of energy.
We also present results using PBE-DFT where we have quantified the BSSE.
In those calculations we use the VTZ basis with added diffuse functions as described in 
Sec.~\ref{wave_functions}. The BSSE corrected Morse fit binding energy and bond length are 
0.79 mHa and 6.45 Bohr. The uncorrected binding energy and bond length are 1.18 mHa and 
6.25 Bohr. Here, the BSSE is 0.39 mHa, roughly half the binding.
In the QMC framework, we do not expect the 
basis incompleteness error to be so important. Although a basis error 
is unavoidable at the VMC level since we use a finite basis set, it is alleviated
by the fact that we fully optimize the AGP and Jastrow bases along with all exponents 
at each $R$. On the other hand, the DMC method for local potentials has no basis error, 
because the stochastic projection to the FN ground state
uses the position representation, which is a complete basis set in configurational space,
and the only bias comes from the FN error. However, in the case of non-local 
pseudopotentials, the localization makes the effective DMC Hamiltonian to depend also on the shape of the
trial wave function (locality error), and so a residual 
error due to incomplete basis could come through. However,
the DMC algorithms used in this work are usually less sensitive to the locality error,
since they are based on a partial localization of the non-local potential,
as described in Sec.~\ref{methods}. Moreover, the locality error, and 
its associated finite basis error, should go
away in the energy differences, because it affects only the core region, which is supposed to be inert,
while in the valence region the dependence on the trial wave function is given just by its nodes.

The good agreement between the VMC and DMC JAGP results, presented
in Subsec.~\ref{jagp_section}, highlights that the basis
set superposition bias is not relevant (smaller than the statistical error of $\sim 0.2$ mHa)
for the fully optimized basis set used in the JAGP wave function,
while the agreement between the projected SJ and JAGP energies, shown in Subsec.~\ref{sj_section},
suggests that the FN bias is negligible.

\subsection{Jastrow correlated Antisymmetric Geminal Power}
\label{jagp_section}

We optimized the variational wave function described in Sec.~\ref{wave_functions} by means of the most recent version
of the SR energy minimization with Hessian acceleration,\cite{sorella:014105} described in Sec.~\ref{methods} and
implemented in TurboRVB.\cite{TurboRVB} The Hamiltonian includes soft pseudopotentials for carbon.\cite{PhysRevB.32.8412}
Although the basis set used here is quite compact,
it turns out that the variational energies are very accurate, as we optimize also the exponents of both the determinantal
and Jastrow part. For instance, the basis set for the hydrogen molecule is a $(2s2p)/[1s1p]$ Gaussian in the AGP expansion,
while it is an uncontracted $(1s1p)$ Gaussian plus a constant in the Jastrow geminal
(a constant generates additional one body terms when multiplied by other orbitals $\chi_{bm}$ in Eq.~\ref{eq:JastrowThreeBody}).
In spite of this small basis set, the  variational energy of an isolated $H_2$ molecule is 
 $-1.174077(29)$, very close the exact result ($-1.174475$).\cite{kolos:404}
The second Gaussian in the $s$ and $p$ contractions of the hydrogen AGP is fairly diffuse, 
their exponents ranging from  $0.05$ to $0.1$, as the distance $R$ 
between  the benzene molecule and the hydrogen dimer shrinks from $15$ to $6$ Bohr.

The basis set of the benzene is slightly larger for the carbon sites ($(6s6p)/[2s2p]$ in the AGP part, uncontracted $(3s 2p)$
in the three-body Jastrow), while for its hydrogen constituents we used just a single $s$ Gaussian
both in the AGP and Jastrow geminals, since they are not supposed to play a key role in the interaction between
the hydrogen molecule and the benzene ring. This fully optimized basis set included in the JAGP wave function gives
a quite good variational energy for aromatic rings.\cite{casula:7110}

We found that the inclusion of the diffuse orbitals in the basis set of the hydrogen molecule
is crucial for the hydrogen-benzene binding, both at the VMC and LRDMC level. On the other hand, some Gaussians
related to the contracted $p$ orbital of the benzene ring become more delocalized in the binding region. This is reasonable, because
the interaction is supposedly driven by the resonance between the carbon $p_z$ and molecular hydrogen $s$ components of the
total wave function. Therefore, the minimal basis set should
include diffuse orbitals on both sides. We would like to stress that the extension of those diffuse orbitals is not determined
\emph{a priori}, but is found by optimizing the wave function with the necessary variational freedom.

After a full optimization of the variational wave function at several distances ($R=5,5.5,6,7,8,10,15$) we carried out
VMC and LRDMC simulations to study the properties of the system, in terms of energetics and charge density distribution.
The LRDMC kinetic parameter in Eq.~\ref{k_operator} which optimizes the lattice space extrapolation is $\mu=3.2$, that allows
one to work with a quite large (and highly efficient) mesh size ($a=0.25$ a.u.). Properly setting the parameters of the
LRDMC effective Hamiltonian is crucial in order to speed up the simulation, and so be able
to resolve the small binding energy of this system.
To check the convergence of our LRDMC energies with respect to the mesh size, we
computed $E(R=6)-E(R=15)$ for $a=0.125$, $0.25$, and $0.5$, as reported in Tab.~\ref{lattice_extrapolation}.
It is apparent that the energy differences are converged within the error bar of $0.25$ mHa
in the lattice space range taken into account. It is therefore accurate to work with $a=0.25$.

\begin{table}[!htp]
\caption{\label{lattice_extrapolation}
LRDMC binding energy ($E(R=6) - E(R=15)$) dependence on mesh size $a$. The energies are reported in mHa,
the lengths are in Bohr.}
\begin{ruledtabular}
\begin{tabular}{l d}
\multicolumn{1}{c}{$a$}
& \multicolumn{1}{c}{$E_{\textrm{binding}}$}
 \\
\hline
0.125 &  1.53(24) \\
0.25  &  1.57(19) \\
0.5   &  2.07(23) \\
\end{tabular}
\end{ruledtabular}
\end{table}

\begin{figure}[!ht]
\centering
\includegraphics[width=0.9\columnwidth]{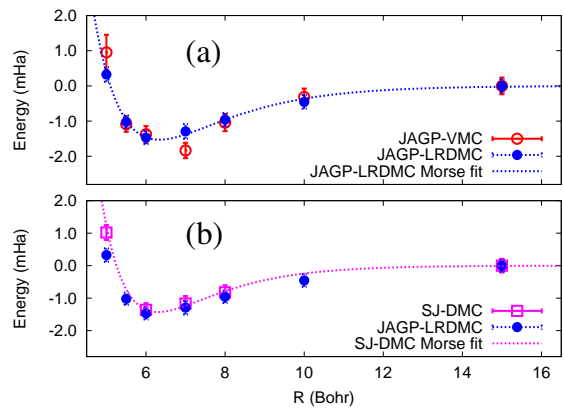}
\caption{(Color online)
Dispersion energy of the hydrogen-benzene bond calculated using various QMC methods for the case where
the hydrogen molecule is oriented vertically with respect to the benzene plane. $R$ is the distance
between the center of mass of the two fragments. The reference for the zero energy difference is taken
at $R=15$. The upper plot compares the variational and the diffusion results using the correlated geminal
wave function, labeled JAGP-VMC and JAGP-LRDMC.  The lower part of the figure compares the best diffusion
results using two types of trial functions, the JAGP (the same as in the upper figure) and the
Slater-Jastrow function labeled as SJ-DMC. The best Morse fits of the diffusion data for the two wave
functions are also plotted as continuous curves. The close agreement of all three results is strong
evidence that the binding curve is accurate and the analytic JAGP variation function (defined in
Eqs.~\ref{eq:JAGPdeterminant}--\ref{eq:JastrowThreeBody}) is a reliable representation of the fully
correlated many-body valence wave function.}
\label{fig:vmc_lrdmc_dmc_multiplot}
\end{figure}

The results of our calculations of the VMC
and LRDMC dispersion curves are presented in Fig.~\ref{fig:vmc_lrdmc_dmc_multiplot}a,
which shows the energy as a function of distance R relative to
the value at  R = 15 for each of the methods.  There
is excellent agreement between the two curves,
with a difference that is less than 0.18 mHa for
most points. Of course, the diffusion calculation leads to
a lower total energy than the variational calculation in
every case, but the agreement of the two methods for the
energy difference supports the idea that our results
are accurate and the calculated binding energy is close
to the exact value.    

In order to extract the values for the equilibrium distance $R_0$ and the binding energy $E_\textrm{b}$,
we fitted our LRDMC points with the Morse function:
\begin{equation}
V(R)=E_\infty + E_\textrm{b}\left[e^{-2a(R-R_0)}-2e^{-a(R-R_0)}\right],
\label{morse_function}
\end{equation}
where $a$ is related to the zero point motion of the effective one dimensional potential $V(R)$,
and $E_\infty$ is chosen to be $E(R=15)$, i.e. the zero of energy. This choice is motivated
by the fact that the overlap of the wave function in between the two fragments is negligible at that distance.
Beyond that point the variation of $V(R)$ up to infinity is much smaller than the statistical accuracy of our points.
We estimated the error on the fitting parameters by carrying out a Bayesian analysis of the fit, in a way similar to what
described in Ref.~\onlinecite{Lucas_K_Wagner_2007}.
Our result is $6.33(15)$ Bohr for the equilibrium distance, and $1.53(12)$ mHa for the
binding energy, as reported in Tab.~\ref{morse}.

\begin{table}[!htp]
\caption{\label{morse}
Fitting parameters of the Morse function (see Eq.~\ref{morse_function}) which minimize the $\chi^2$ of the
JAGP-LRDMC and SJ-DMC data sets.
Their error is computed by means of a Bayesian analysis\cite{Lucas_K_Wagner_2007}
based on the statistical distribution of the FN energy points.
The energies are reported in mHa, the lengths are in Bohr.}
\begin{ruledtabular}
\begin{tabular}{l d d}
& \multicolumn{1}{c}{JAGP} & \multicolumn{1}{c}{SJ} 
 \\
\hline
$a$                  &   0.56(7)      &   0.66(9)        \\ 
$E_\textrm{b}$  &   1.53(12)     &   1.43(16)       \\ 
$R_0$                &   6.33(15)     &   6.31(21)        \\ 
\end{tabular}
\end{ruledtabular}
\end{table}

\subsection{Slater-Jastrow Trial Function}
\label{sj_section}

At this point, it is interesting to make a comparison with a simple SJ wave function to
determine whether the use of the JAGP is necessary to get the correct dispersion energy
out of the FN projection. We generated the Slater part by performing DFT calculations with the PBE
functional using the VTZ basis\cite{burkatzki:234105} with diffuse functions from the aug-cc-pVTZ
basis.\cite{kendall:6796} All the DFT calculations were done using 
Gaussian03.\cite{g03} We chose to use a very simple Jastrow factor because
our goal was to improve DMC efficiency as opposed
to obtaining a well converged binding curve at the VMC level.
Therefore our Jastrow includes only one- and two-body contributions
with three non-cusp terms in the expansion (see
Eqs.~\ref{eq:wmjastrow1} and \ref{eq:wmjastrow2}). Furthermore, a single cusp term is added in each of the
hydrogen and electron-electron Jastrow factors. No cusp is included for carbon due to the use
of a soft pseudopotential.\cite{PhysRevB.32.8412}

\begin{table}[!htp]
\caption{\label{SJtimestep}
DMC binding energy ($E(R=6) - E(R=15)$) dependence on time step $\tau$.
The energy extrapolated for $\tau \rightarrow 0$ is within one error bar
from the point at $\tau=0.01$. Therefore, we chose $\tau=0.01$ as the time step
for all our DMC simulations. The energies are reported in mHa, the time steps are in Ha$^{-1}$.}
\begin{ruledtabular}
\begin{tabular}{l d}
\multicolumn{1}{c}{$\tau$} & \multicolumn{1}{c}{$E_{\textrm{b}}$} \\
\hline
0.01  &  1.38(19) \\
0.02  &  0.93(19) \\
0.04  &  0.64(15) \\
\end{tabular}
\end{ruledtabular}
\end{table}

All the Monte Carlo calculations with the SJ wave function
were done using QMCPACK.\cite{qmcpack} The Jastrow factor was
optimized within the VMC framework using the conjugate gradient 
method,\cite{HestenesStiefel1952} as explained in Sec.~\ref{methods}.
While the SJ variational energy is quite poor, its quality is not directly reflective of the DMC energy, which
depends only on the nodes of the trial wave function. In this case, indeed, we found that the DFT nodes are very good
by carrying out DMC simulations with the non-local scheme described in Sec.~\ref{methods}.
Our projection was done in time steps of $\tau = 0.01$ which we found to be converged as reported in Tab.~\ref{SJtimestep}.
Remarkably, the DMC-SJ energies are in very good  agreement
with the LRDMC-JAGP data points (see Fig.~\ref{fig:vmc_lrdmc_dmc_multiplot}b).
Indeed, the SJ fitting parameters of the Morse dispersion curve (Eq.~\ref{morse_function}),
such as binding energy, equilibrium distance, and curvature, differ from the
JAGP ones by less than one error bar (Tab.~\ref{morse}).
This consistency between different trial wave functions signals that the FN bias is negligible
and the results are well converged. Moreover, in addition to the nodes of the PBE wave function being good,
the PBE binding energy is underestimated only by a factor 2 with respect to our best value.
It is notable that the PBE functional performs quite well,
even though it does not include any VdW contribution.
In the case of a pure VdW bond, the PBE result should be much poorer, as already
pointed out by Hamel and C\^ot\'e.\cite{HamelCote2004}
This is suggestive of a more complex binding mechanism which goes beyond the standard physisorption.
We will focus on this point in Sec.~\ref{discussion}.

\begin{figure}[!ht]
\centering
\includegraphics[width=0.9\columnwidth]{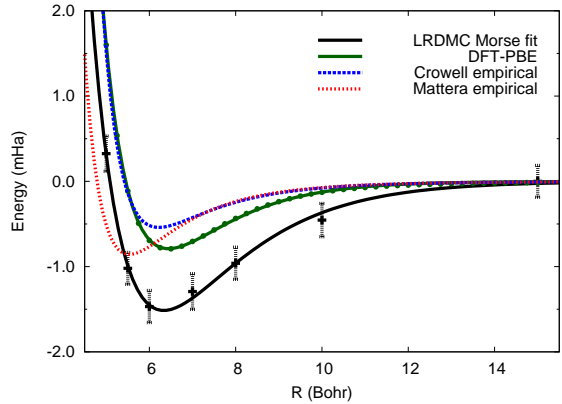}
\caption{(Color online)
Results of different theoretical descriptions of hydrogen-benzene binding as a function of intermolecular
distance $R$ where hydrogen is situated perpendicular to benzene. The solid black line and associated error
bars show the JAGP-LRDMC data and Morse fit (zero binding energy is taken at $R=15$ Bohr). 
The green curve shows the PBE-DFT counterpoise corrected result 
using the VTZ basis\cite{burkatzki:234105}
supplemented by the diffuse functions from the aug-cc-pVTZ basis.\cite{kendall:6796} The dotted blue curve
(shallowest) is the empirical potential devised by Crowell and Brown\cite{crowell_1982} that takes into
account the bond asymmetry of the $sp^2$ hybridized carbon atom. The
dotted red line is the empirical potential by Mattera \textit{et al.}\cite{mattera_1980} that seeks to
reproduce the hydrogen bound states over graphite by a much simpler model.
The Mattera potential does not take into account bond asymmetry.}
\label{fig:morse_empirical_pbe_mp2}
\end{figure}

\section{Comparison to other work}
\label{otherwork}

The hydrogen-benzene system has been the subject of several theoretical works, whereas
to our knowledge
no direct study of this system has been carried out on the experimental side.
Hydrogen adsorbed on metal-organic frameworks (MOF),
where benzene-like structures serve as ligands, has been studied
by Rosi \textit{et al.}\cite{Rosi05162003} who performed inelastic neutron scattering (INS) measurements.
The INS data could be related to the rotational states of hydrogen 
adsorbed over benzene. However, the binding sites in the MOF structure are not known with certainty, and thus
it is hard to find a one-to-one correspondence between the experiment
and the isolated hydrogen-benzene compound.

Given the lack of direct experimental data for this system, we
compare our results with those from empirical models that are often used to estimate complex system properties,
such as the hydrogen storage capabilities of carbon nanotubes
and fullerene nanocages.\cite{SurfSci8.767,wang:577} Here we consider two empirical models, both
derived from experiments of hydrogen molecules scattered on graphite surface,
carried out by Mattera \textit{et al.}.\cite{mattera_1980}
To reproduce their data, they proposed
a simple model interaction between the carbon atoms and the hydrogen dimer which
depends only on the distance from the graphite layers by
assuming lateral average. This model was improved later by
Crowell and Brown,\cite{crowell_1982} who constructed an empirical potential 
based not only on the experimental scattering data
but also on the polarization constants built in the VdW (6,12) potential.
Their model assumes both a radial and angular dependence, which takes into account
the $sp^2$ hybridization asymmetry of carbon atoms in graphitic and aromatic compounds.
We applied these potentials to the hydrogen-benzene system by summing the terms for the 6 carbons taking into
account distance and, for the Crowell potential, the angle the hydrogen-carbon interaction makes with the benzene
$C_6$ axis. Both empirical potentials significantly underbind the system, roughly by factors
of 3 and 2 respectively when compared to the JAGP LRDMC results (see Fig. \ref{fig:morse_empirical_pbe_mp2}).
More precisely, Mattera's interaction gives a binding energy of 0.86 mHa at 5.6 Bohr, while
Crowell's gives a minimum of 0.54 mHa at 6.2 Bohr.

Hamel and C\^{o}t\'{e}\cite{HamelCote2004} calculated the dispersion curves using DFT with the
local density and generalized gradient approximations (LDA and GGA) where the GGA is implemented in the
PBE density functional.\cite{PhysRevLett.77.3865,PhysRevLett.78.1396} Their calculations used a plane wave
basis with a 60 Ha cutoff. They found that the DFT-LDA gives the strongest binding (3.30 mHa),
while the DFT-PBE binding is much weaker (0.69 mHa). 
This is consistent with the general overbinding of LDA and underbinding of PBE.
It is also well known that DFT is not a favorable method for systems where van der Waals forces play an important
role;\footnote{Note that in our work, we used the single-body orbitals from the PBE-DFT calculation in the
Slater-Jastrow wave function. The DMC energies depend only on the accuracy of the nodes of the many-body wave function.
The DMC calculation includes van der Waals attraction and other terms and the result is independent of the errors in
the PBE functional for the energy.}
in those cases, MP2 and CCSD(T) can be applied with more reliability.
Hamel and C\^{o}t\'{e} also calculated binding curves using those theories. They found MP2/6-311+G(2df,2p)
binding of 1.58 mHa and CCSD(T)/6-31+G(d,p) binding of 0.65 mHa.

Perhaps the most careful and accurate MP2 and CCSD(T) calculations were done by H\"{u}bner
\textit{et al.}\cite{Hubner2004}
In order to resolve the weak interaction between hydrogen and benzene, high accuracy is required, and so
a large basis set is needed to reduce both basis set superposition and incompleteness errors
which are a significant fractions of the binding energy 
(the BSSE was found to be as much as $\sim 25\%$ of the final estimated binding).
On the other hand, the use of a larger basis set is limited by a poorer scaling of the calculations, particularly at
the CCSD(T) level of theory, which is the most expensive. 
In their work, H\"{u}bner \textit{et al.} optimized the binding distance using MP2 with the TZVPP basis.
They found a center-of-mass distance of 5.80 Bohr and a binding energy of 1.47 mHa. This geometry
was then used for further MP2 and CCSD(T) calculations.
The CCSD(T) method with the same TZVPP basis gives 1.17 mHa,
while the MP2 theory was pushed up to a aug-cc-pVQZ$^\prime$ basis to
give a binding of 1.83 mHa, a significant increase from the TZVPP basis.
At this point, it is possible to estimate the true binding energy 
by correcting the best MP2 energy with the CCSD(T)-MP2 difference obtained at the TZVPP level.
This gives a value of $\sim 1.5$ mHa, remarkably close to the JAGP LRDMC binding of 1.53 $\pm$ 0.12 mHa,
found in this work.

\section{Analysis of the bonding}
\label{discussion}

\begin{figure}[!ht] 
\centering
\includegraphics[width=0.7\columnwidth]{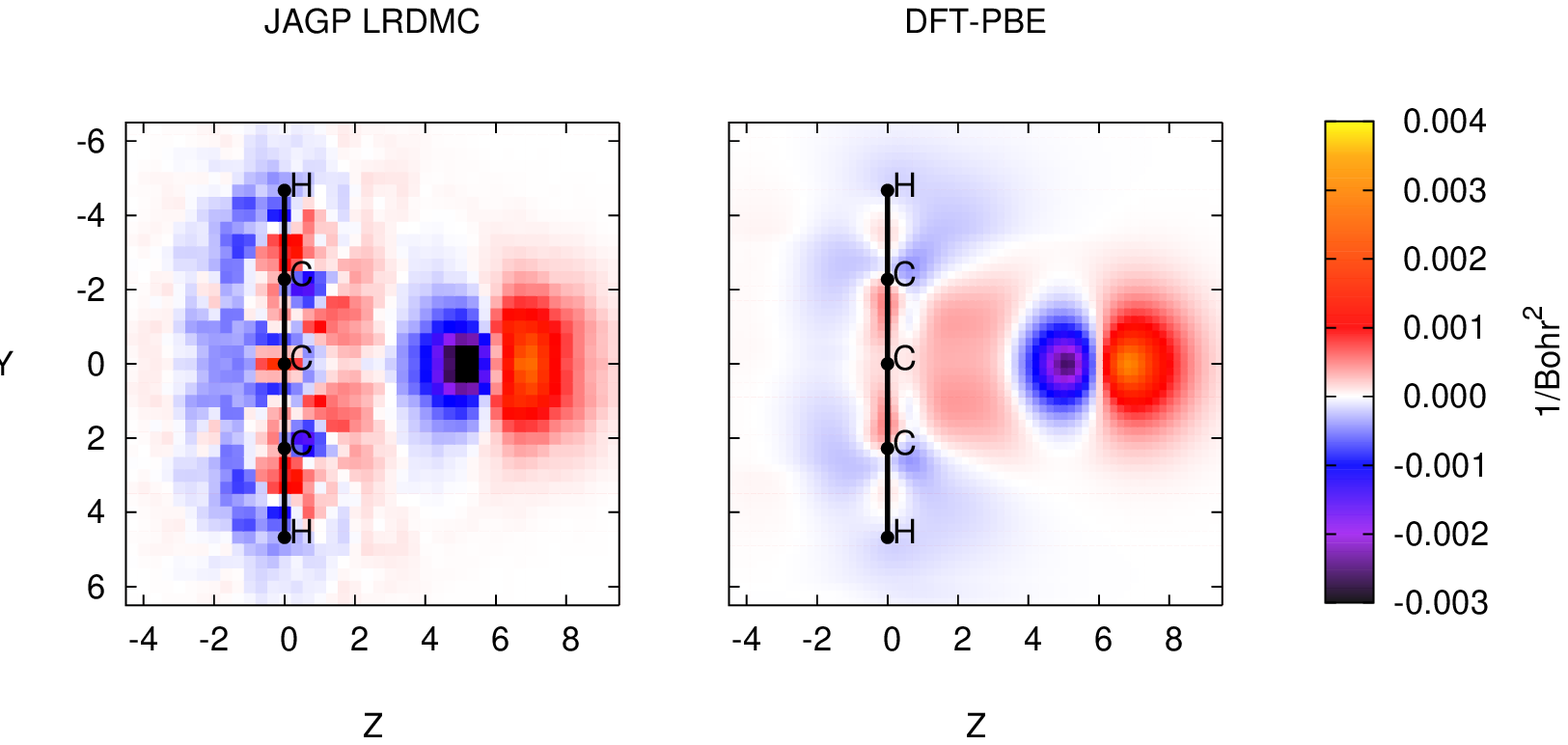}
\caption{(Color online)
Contour plots of the difference in projected electronic charge per unit area between hydrogen-benzene
separated by $6$ Bohr and the isolated hydrogen and benzene using JAGP-LRDMC and PBE-DFT. This plot shows how the charge per unit area
changes as hydrogen and benzene interact, the x-axis have been integrated over so that the charge per unit area has been
projected into the yz-plane. (Left) The areal charge density difference is a mixed estimate of LRDMC calculations with
a JAGP trial wave function. (Right) computation is done within the PBE-DFT framework using the 
VTZ basis\cite{burkatzki:234105} supplemented by diffuse functions from the aug-cc-pVTZ basis.\cite{kendall:6796} 
The combined hydrogen-benzene system basis was also used in the isolated benzene and hydrogen calculations.}
\label{fig:multiplot_density}
\end{figure}

\begin{figure}[!ht] 
\centering
\includegraphics[width=0.9\columnwidth]{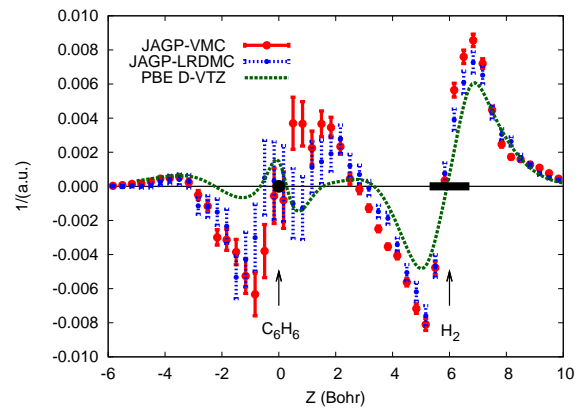}
\caption{(Color online)
Plot of the difference in linear electronic charge density between hydrogen-benzene separated by $6$ Bohr and the
isolated hydrogen and benzene using three theories. This plot shows how the linear charge density
distribution changes as hydrogen and benzene interact, the x- and y-axes have been integrated over so that only
the z-axis is shown. The positions of the hydrogen and benzene are indicated in the graph. The solid red data
with error bars show the induced charge changes using the analytic JAGP wave function at the VMC level.
The dotted blue data with error bars show the mixed estimate of the density,
given by the LRDMC projection of the JAGP trial wave function. The dotted green line without error bars shows the PBE-DFT 
charge density difference using the VTZ basis\cite{burkatzki:234105} supplemented by the diffuse functions from
the aug-cc-pVTZ basis.\cite{kendall:6796} 
The combined hydrogen-benzene system basis was also used in the isolated benzene and hydrogen calculations.}
\label{fig:rho_z_vmc_dmc_pbe}
\end{figure}

In order to investigate more deeply the physics of hydrogen adsorbed 
on benzene, we study the induced difference in electronic density at the equilibrium bond 
distance with respect to the separated fragments.
For this study we compare our best DMC results to the density functional calculation using
the PBE functional. The QMC densities are calculated from the optimized correlated geminal (JAGP)
as a mixed estimator, which is an accurate representation of the DMC results since the diffusion calculation leads
to only small changes (within the error bar)
from the VMC density. The contour plot in Fig.~\ref{fig:multiplot_density} shows the
difference in the calculated electron density at the separation $R=6$ Bohr. Here, the electron density of the isolated
molecules has been subtracted from the combined system so that the change in charge distribution
due to bonding is apparent. In this figure the benzene ring lies in the $xy$ plane at $z=0$ and
the hydrogen molecule is oriented along the $z$ axis, with its center of mass at $z=6$ Bohr.
The two dimensional plot in the $yz$ plane is generated by integrating the density distribution over the $x$
coordinate. As one can see, the hydrogen molecule is polarized by the electronic repulsion
with the benzene cloud, which pushes the electrons to the opposite side of the molecule,
leading to a static dipole moment on the hydrogen. On the other hand, the density redistribution in the
benzene is non trivial, and shows patches of charge accumulation and depletion. To catch the net effect
of this redistribution, we integrated the density also over the $y$ coordinate, and obtained an effective
linear density profile, plotted in Fig.~\ref{fig:rho_z_vmc_dmc_pbe}. Here, it is apparent that the overall effect
on the benzene is the formation of another effective dipole moment, oriented to the same direction as
the static dipole moment on the hydrogen molecule, which lowers the electrostatic energy.
Notice that in Fig.~\ref{fig:rho_z_vmc_dmc_pbe} we have plotted separately the VMC and the LRDMC mixed estimate
of the densities. The close agreement supports our conclusion the VMC wave function
is very accurate not only for the energy but also for other properties such as the density.

At large distances the attractive interaction is due to VdW dispersive forces, which is
included in the Monte Carlo calculations.  At short distances the interaction is repulsive due to overlap
of the closed shells, which would lead to density displaced outward on both the hydrogen
and benzene, i.e. opposite dipoles on the two molecules.
However, Figs.~\ref{fig:multiplot_density} and \ref{fig:rho_z_vmc_dmc_pbe} show
that the hydrogen-benzene bond is not a pure VdW interaction, since in
the binding region also electrostatic effects come in
with the onset of dipolar interactions that lower the charge repulsion.
For comparison, density differences calculated using the PBE density functional are also shown in
Figs.~\ref{fig:multiplot_density} and \ref{fig:rho_z_vmc_dmc_pbe} at the separation $R=6$ Bohr.
Of course, the PBE functional does not include VdW interactions so that the binding
decreases too rapidly at large distance as shown in Fig.~\ref{fig:morse_empirical_pbe_mp2}.
Nevertheless, near the equilibrium distance the density is similar to the QMC result but with
smaller magnitude of the change in density, which is consistent with the fact that
the PBE functional underbinds the system.
It is well known that GGA functionals like PBE tend to underbind because
they favor systems with larger gradients, whereas LDA
tends to overbind molecules and solids since it favors more homogeneous systems.\cite{Martin2004}
Recent work by Langreth \textit{et al.}\cite{PhysRevLett.92.246401,thonhauser:125112} has led to improved functionals
including van der Waals interactions; however, they have not been considered here.

\section{Summary and Conclusions}
\label{summary}

We have presented VMC and DMC results for the 
adsorption of hydrogen on a benzene ring 
and compared them with previous work. We used two types of variational correlated wave functions,
a SJ function with DFT-PBE optimized single-body orbitals and a JAGP function fully optimized
at the VMC level by means of the SR energy minimization.
In this work, we have shown strong evidence that
our results are very accurate since we have found essentially the same results
in three independent QMC calculations: one JAGP-VMC variational simulation
with no FN error, and two DMC simulations based on different trial wave functions
(JAGP and SJ) with possibly different nodes and no basis set errors.
The agreement among our three calculations is within $\simeq 0.2$ mHa,
which is mainly due to statistical accuracy on the QMC energies, and
gives an upper bound for the magnitude of underlying errors, such as
the basis set incompleteness and the FN bias.

Our best estimate for the binding energy is 1.53(12) mHa at an equilibrium distance
of 6.33(15) Bohr, obtained by using the LRDMC method with the nodes of the JAGP
wave function. 
Our result agrees well with the conclusion of
H\"{u}bner \textit{et al.}\cite{Hubner2004} who used MP2 and CCSD(T) methods,
and estimated the binding to be $\sim 1.5$ mHa based on extrapolation which accounts for
basis set and level of theory.
The resulting binding energy is 2--3 times larger
than those given by empirical potentials\cite{crowell_1982,mattera_1980} and DFT-PBE
calculations which are often employed in more complex systems, suggesting that
their results could be substantially affected by this lack of accuracy.
It would be interesting to extend the present work, by studying
the transferability of such empirical potentials on other aromatic and graphitic structures.

We proved that the JAGP wave function provides a very accurate dispersion curve
for this system already at the variational level.
This result is remarkable, because we were able to derive a compact analytic form
which can be used for accurate determination of properties other than the energy
by means of the VMC method with no sign problem. The JAGP wave function
captures the resonating valence bonds of benzene in
its geminal construction as well as the van der Waals interaction through
many-body correlations in the Jastrow factor,
as shown in previous work on benzene dimer.\cite{sorella:014105}
The basis for both the AGP and Jastrow geminals is of compact Gaussian form that
does not go beyond $p$-orbitals, 
but includes diffuse orbitals with optimized exponents.

By means of the DMC method, we also studied the hydrogen-benzene problem
using a more conventional SJ wave function.
The single-body orbitals included in the Slater determinant
were derived using the GGA PBE density functional with a VTZ Gaussian basis\cite{burkatzki:234105}
modified to include diffuse functions from the aug-cc-pVTZ basis,\cite{kendall:6796} 
as discussed in Subsec.~\ref{wave_functions}, which
is essential for an unbiased DMC sampling of long-range VdW effects.
The Slater basis set is roughly 4 times larger than its JAGP counterpart 
whereas the Jastrow factor is of minimal form, satisfying cusp conditions and
improving computational scaling.
Our findings suggest that for this particular problem the geminal form is not essential
to get an accurate DMC energy, and can be replaced with a Slater determinant
and DFT optimized orbitals in the DMC calculations.
While the JAGP uses a more compact basis, the SR optimization involves a large number of parameters
coming from the Jastrow and AGP geminals expanded on atomic orbitals.
This makes the SJ wave function with DFT-PBE single-body orbitals more desirable for DMC calculations
in larger related systems.

Finally, we examined how
the electronic density of the isolated molecules changes in the bond region.
The change in density, displayed in Figs.~\ref{fig:multiplot_density} and~\ref{fig:rho_z_vmc_dmc_pbe},
shows that near the equilibrium distance there is the formation of static dipoles that
can lower the electrostatic energy, indicating a bonding mechanism beyond VdW.
Density functional calculations using the PBE functional lead to similar density profiles but
with smaller magnitude, in agreement with the well known underbinding tendency 
of that functional. This means that the interaction between hydrogen and benzene 
is not a pure VdW effect, since it can be partially captured by a DFT-PBE formalism
which does not include dispersive interactions. This also clarifies why the
DFT-PBE nodes of the SJ wave functions are very good, and equivalent to the
JAGP nodes to predict the correct binding energy at the DMC level.

To conclude, we have reported on a detailed analysis of the hydrogen 
adsorption over molecular benzene by QMC methods, which are shown to be very accurate and reliable to
predict the energetics and other physical properties of the system. This framework is therefore promising
to study hydrogen interacting with graphitic or other aromatic compounds,
particularly important for the hydrogen storage problem.

\acknowledgments

We are indebted to S. Chiesa for discussions in the early stages of this work and to S. Hamel
for fruitful discussions on hydrogen storage and calculations on the hydrogen-benzene system.
We thank Claudia Filippi for illuminating suggestions.
Calculations were done at the National Center for Supercomputing Applications (NCSA)
and Computational Science and Engineering (CSE) facilities at Illinois and resources 
of the National Center for Computational Sciences at Oak Ridge National Laboratory, 
which is supported by the Office of Science of the U.S. Department of Energy under Contract No. DE-AC05-00OR22725.
This work was partially supported by the NSF grant DMR-0325939 and U.S. Army A6062 UMC00005071-3.
One of us (M.C.)  acknowledges support in the form of the NSF grant DMR-0404853.

\bibliography{bibliographic_database}
\end{document}